\documentclass[prb,reprint,amsmath,amssymb,aps,superscriptaddress,longbibliography]{revtex4-1}
\usepackage{float} 
\usepackage{graphicx}
\usepackage{dcolumn}
\usepackage{bm}
\usepackage[unicode]{hyperref}
\hypersetup{
   unicode=true,          
   plainpages=false,
   colorlinks=true,       
   linkcolor=red,          
   citecolor=blue,        
}

\usepackage{times}
\usepackage{tikz}
\usetikzlibrary{calc}
\usepackage{todonotes}
\usepackage{physics}
\usepackage{cancel}
\usepackage{xinttools}

\usepackage{svg}
\usepackage{wasysym} 

\graphicspath{{fig/},{python_fig/}}

\newcommand{\rmi}{\mathrm{i}\,}

\definecolor{mylightblue}{RGB}{37, 148, 212}
\definecolor{myred}{RGB}{200, 70, 48}
\definecolor{mydarkblue}{RGB}{9, 64, 116}
\definecolor{mygreen}{RGB}{0, 155, 114}
\definecolor{myyellow}{RGB}{255, 221, 74}
\definecolor{myyellowdark}{RGB}{180, 145, 0}

\newcommand{\ysym}[1]{%
\begin{tikzpicture}[#1]%
\draw (0,0) -- (0,1);%
\draw (0,1) -- (1,2);%
\draw (0,1) -- (-1,2);%
\end{tikzpicture}%
}

\newcommand{\lambdasym}[1]{%
\begin{tikzpicture}[#1]%
\draw (0,2) -- (0,1);%
\draw (0,1) -- (1,0);%
\draw (0,1) -- (-1,0);%
\end{tikzpicture}%
}

\begin{document}
\title{Majorana-XYZ subsystem code}
\author{Tobias Busse}
\email{research@tobiasbusse.de}
\author{Lauri Toikka}
\email{lauri.toikka@gmail.com}

\date{\today}
\begin{abstract}
Building fault-tolerant quantum computers is a current priority for the quantum computing industry and academic counterparts. We present a quantum error correction code encoding a single logical qubit into $n=L^2$ physical qubits with distance $d = L$ as a subsystem code, and $n-O(\sqrt{n})$ logical qubits with distance $d=3$ as a standard stabiliser code. The logical single-qubit gates in the subsystem code are transversal. Under independent $X,Z$ noise we find a threshold of $1.4\% \pm 0.1\%$ for odd distance and $1.8\% \pm 0.2\%$ for even distance for the subsystem variant. The code detects every 1- and 2-qubit error, and additionally every error of weight 3 and higher (constrained by the distance) that is not a product of the 3-qubit check operations. These operations act on the gauge qubits leaving the subsystem code space invariant. The code can be implemented by qubits in a triangular configuration with the plaquettes as 3-qubit check operations.

\end{abstract}
\maketitle

\section{Introduction}

In order to build a quantum computer capable of demonstrating any advantage over classical computers, we either need to sufficiently suppress the physical errors in the qubits, or execute the computation within a quantum error-correction (QEC) scheme~\cite{PRXQuantum.2.040101,Gottesman2014}. It is not possible to completely suppress all errors, but with QEC it suffices if the physical error rate is below the code threshold. In general, the smaller the error rate the larger the quantum computation we can run with given resources. Modern QEC schemes can tolerate physical errors of around $10^{-4}$ per gate~\cite{Knill2005}, though thresholds upto $1\%$ have been reported~\cite{PhysRevA.86.032324,PhysRevA.83.020302,Raussendorf_2007}. It's an active research front to on one hand suppress physical error rates to be commensurate with the threshold of QEC, and on the other hand to find better theoretical QEC protocols with lower overhead~\cite{aliferis2005quantumaccuracythresholdconcatenated}. 

We present a subsystem QEC protocol that relies on nearest-neighbour qubit coupling and local weight-3 check measurements. The code encodes a single logical qubit and the logical operators are homologically non-trivial cycles. We find thresholds of $p_\mathrm{threshold} = 1.4\% \pm 0.1\%$ for odd distance and $p_\mathrm{threshold} = 1.8\% \pm 0.2\%$ for even distance. In the case of unbiased noise or no multi-qubit measurements, these thresholds compare favourably with those reported for the Floquet honeycomb code under periodic boundary conditions and surface code in Ref.~\cite{Gidney2021faulttolerant} ($0.2-0.7\%$), those reported for subsystem codes in Ref.~\cite{PhysRevX.11.031039} ($0.67-0.81\%$), and the improved thresholds for the Floquet honeycomb and 4.8.8. Floquet code reported in Ref.~\cite{PRXQuantum.4.010310} ($1.0-1.3\%$). Subsystem codes have had lower thresholds compared to the surface code~\cite{PhysRevX.11.031039}. The thresholds here are almost an order of magnitude improvement of the subsystem thresholds found in Ref.~\cite{Brown2016-dq} ($0.31\%$) and approaching the optimal threshold for the toric code under the same error model~\cite{Wang2003-zy} ($2.9\%$).

While Kitaev's original honeycomb code~\cite{Kitaev2006} is promising as a subsystem code with 2-qubit measurements, there can only be protected logical qubits under a dynamical Floquet measurement schedule construction~\cite{Lee_2017,Hastings2021}. Here, we report a subsystem code with 3-qubit measurements and state-of-the-art threshold that natively contains a single logical qubit without the need for Floquet measurements. Given the improvement in the threshold of the Kitaev honeycomb code under a circuit model with native two-body measurements~\cite{Gidney2021faulttolerant}, we expect the threshold to improve further using the native operations of Majorana-based architectures~\cite{2025MSFT_Majorana-1,PRXQuantum.4.010310}.

\begin{figure}[t]
        \centering
\begin{tikzpicture}[scale=0.6]

	\begin{scope}[yshift=0,every node/.append style={
    	    yslant=0.5,xslant=-1},yslant=0.5,xslant=-1]
    	\fill[blue!10] (0,0) rectangle (8,6);
    	\draw[blue!5,very thick] (0,0) rectangle (8,6);
            	    		
        \clip (0,0) rectangle (8,6);
    	\begin{scope}[y=(60:1),shift={(1,0)}]
       		\foreach \x in {-8,-6,...,8}{
  				\draw[help lines,dashed]
    				(\x,-8) -- (\x,8)
   					(-8,\x) -- (8,\x) 
    				[rotate=60] (\x,-8) -- (\x,8) ;}
    				
	    		\draw [line width=0.4mm, draw=black, opacity=1.0]
	       			(2,2) 	       -- (4,2) 	        --  (2,4)	        -- cycle;
	        
	        	\draw [line width=0.4mm, draw=black, opacity=1.0]
	       			(-2,6)    -- (0,6) --  (0,4) 	       -- cycle;
			    
    	\end{scope}    		
	\end{scope}

	\begin{scope}[yshift=0,every node/.append style={
    	    yslant=0.5,xslant=-1},yslant=0.5,xslant=-1,
    	    scale=2/sqrt(3),
  				majorana/.style={circle,shading=ball,minimum size=0.25,ball color=orange},
  				majoranablue/.style={circle,shading=ball,minimum size=0.25,ball color=blue},
  				majoranamag/.style={circle,shading=ball,minimum size=0.25,ball color=magenta}	]				
 			\clip (0,0) rectangle (${(sqrt(3)/2)}*(8,6)$);
			\begin{scope}[rotate=30]
			  	\foreach \x in {0,...,6} {
			    	\coordinate(X) at (0:1.5*\x);
			    	\ifodd\x
				      \def\ymax{5}
				      \coordinate(X) at ($(X)+(0:0.5)+(-120:1)$);
				    \else
				      \def\ymax{4}
				    \fi
				    \foreach \y in {-\ymax,...,\ymax} {
				      \coordinate (\x-\y) at ($(X)+(60:\y)+(120:\y)$);
				      \draw[orange,thick] (\x-\y) +(-60:1)
				      \foreach \z [remember=\z as \lastz (initially 5)] in {0,...,5} {
				        -- coordinate(\x-\y-\lastz-m) +(\z*60:1) coordinate(\x-\y-\z)
				      } -- cycle;
				    } 
				  }
			\end{scope}
			  
				\foreach \dt in {1-1,2-0,3-0,4-1,1-2,2-1,3-1,4-0,2-2,3-2,4-1,5-1,2--1} {
		  		\foreach \z in {0,...,5} {\node[majorana] at (\dt-\z) {};}}
		  		\foreach \dt in {1-1-5,1-1-0,2-1-5,2-1-3} {\node[majoranablue] at (\dt) {};}
				  \draw[thick,blue] (1-1-0) --  (1-1-5); 
				  \draw[thick,blue] (1-1-0) --  (2-1-5); 
				  \draw[thick,blue] (1-1-0) --  (2-1-3); 			

		  		\foreach \dt in {2-0-5,2-0-0,2-0-4,3-0-4} {\node[majoranamag] at (\dt) {};}
				  \draw[thick,magenta] (2-0-5) --  (2-0-0); 
				  \draw[thick,magenta] (2-0-5) --  (2-0-4); 
				  \draw[thick,magenta] (2-0-5) --  (3-0-4); 	

							
	\end{scope}
    		
    \draw[-latex,thick](-1.5,0)node[left]
        {$\mathsf{Majorana \; fermion}$} to[out=0,in=180] (0.0,0.8);

    \draw[-latex,thick](5.5,0.0)node[left]
        {$\mathsf{Topological \; order}$} to[out=50,in=-30] (4.0,2.1);
        

    \begin{scope}[
    	yshift=100,every node/.append style={
    	    yslant=0.5,xslant=-1},yslant=0.5,xslant=-1]
        \filldraw [white,fill opacity=.8] (0,0) rectangle (8,6);
        \draw[black,very thick] (0,0) rectangle (8,6);
        
        \clip (0,0) rectangle (8,6);
    	\begin{scope}[y=(60:1),shift={(1,0)}]
       		\foreach \x in {-8,-6,...,8}{
  				\draw[help lines,dashed]
    				(\x,-8) -- (\x,8)
   					(-8,\x) -- (8,\x) 
    				[rotate=60] (\x,-8) -- (\x,8) ;
    				
			\foreach \y in {-8,-6,...,8}{
				\coordinate  (Dot\x\y) at (\x,\y);
    			\node[draw,circle,inner sep=2pt,fill] at (\x,\y) {};
    			
				\draw[ultra thick,-latex,cyan,shift=(Dot\x\y.center),
    				rotate around={rand*180:(Dot\x\y.center)}] (-0.5,0) -- (0.5,0);
    			}}
 	
    		\draw (Dot26)+(0,-1mm) node[anchor=north,cyan]{Spin-$\frac{1}{2}$};
    			
    		\draw [line width=0.4mm, draw=black, opacity=1.0]
       			(2,2) node[draw,circle,inner sep=1pt,fill=white,line width=0.21mm]{\large $Y$}
       -- (4,2) node[draw,circle,inner sep=1pt,fill=white,line width=0.21mm]{\large $X$}
        --  (2,4) node[draw,circle,inner sep=1pt,fill=white,line width=0.21mm]{\large $Z$} 
        -- cycle;
        
        	\draw [line width=0.4mm, draw=black, opacity=1.0]
       			(-2,6)  node[draw,circle,inner sep=1pt,fill=white,line width=0.21mm]{\large $X$}
       -- (0,6) node[draw,circle,inner sep=1pt,fill=white,line width=0.21mm]{\large $Y$}
       --  (0,4)  node[draw,circle,inner sep=1pt,fill=white,line width=0.21mm]{\large $Z$}
       -- cycle;
		\node at (barycentric cs:Dot22=1,Dot42=1,Dot24=1) {\large $\hat{T}^\Delta$}; 
		\node at (barycentric cs:Dot-26=1,Dot06=1,Dot04=1) {\large $\hat{T}^\nabla$};    
    
		\end{scope}
    \end{scope}
\end{tikzpicture}        
        
        \caption{\label{fig:model}The Majorana-XYZ code. The physical system is a honeycomb arrangement of Majorana fermions, for example, quantised vortices in an $s$-wave superfluid in the topological phase~\cite{toikka_2019}. Top: the representation in terms of an equivalent spin-$\frac{1}{2}$ model on a triangular lattice. The four-Majorana interaction terms mapping to the Hamiltonian down and up triangles in blue and magenta respectively. The system is highly frustrated, where the origin of frustration comes from the anti-commutation of any two up-up or down-down corner-sharing Hamiltonian triangles. The code space used to store and process quantum information is a specific symmetry sector of the physical topological quantum spin liquid like state.}               
    \end{figure}

The code is a specific symmetry sector of the physical system defined by Majorana fermions in a honeycomb arrangement with only nearest-neighbour Majorana-Majorana interactions. The honeycomb Majorana Hubbard model was originally studied by Li and Franz~\cite{Majorana_Hubbard_Honeycomb}. This system is equivalent to spin-$\frac{1}{2}$ particles on a triangular lattice where the interactions are 3-spin operators acting at the vertices of the triangular plaquettes (Fig.~\ref{fig:model} and Sec.~\ref{sec:physsyst}). We study elsewhere~\cite{BTL} the nature of the quantum and classical ground state of the fragmented Hilbert space of this Majorana-XYZ Hamiltonian that results from the rich subsystem symmetries. 

Experimentally, Majorana fermions are postulated to appear in vortex cores in a topological superfluid consisting of ultra-cold fermionic atoms~\cite{toikka_2019,zhou2011,volovik1999,simonucci15}, where the non-Abelian Majorana braiding and fusion dynamics translates into superfluid vortex dynamics and can therefore arise in a natural way. Commercial efforts by Microsoft Research have focussed on creating Majoranas in a solid-state device termed the Majorana-1 and Majorana-2 chips~\cite{2025MSFT_Majorana-1,aghaee202620secondparitylifetime} where braiding is achieved with in-situ interrogation. The Majorana-based quantum technology promises to get millions of qubits on a single chip~\cite{2025MSFT_Majorana-1} or vortex lattice~\cite{zwierlein2005,toikka_2019} in a superfluid drop.
From the point of view of implementation in practice, having to do only low-weight check measurements is a particularly attractive feature for a code. This is the driving idea behind subsystem codes~\cite{PhysRevA.73.012340}, where the geometrical locality of check measurements is the reward for sacrificing Hilbert space to non-encoding gauge degrees of freedom. The fewer qubits that we need to involve per error correction measurement the lower the probability of introducing additional errors resulting in a lighter circuit implementing the error correction cycle and thus reducing the overhead of the QEC protocol. Most hardware approaches are limited to nearest-neighbour connectivity meaning geometrically-local low-weight check measurements are particularly desirable.

The Majorana-XYZ code is a surface code where the stabiliser generators are not local, which is typically taken as a prerequisite for topological codes~\cite{PhysRevA.81.032301,Dennis2002,Kitaev1997,Kitaev2003}. The stabilisers however do not constitute the actual error correction measurements. Instead, the physical checks correspond to the gauge operators, whose products give the stabiliser syndrome. By topological protection we mean here that the logical information cannot be modified locally by an undetectable (or detectable) error: action by the gauge group, which constitutes the undetectable errors of the code, does not damage the encoded logical information; even if undetectable by the code, they only act on the gauge qubits.

\section{\label{sec:physsyst}Majorana-XYZ code}
The Majorana-XYZ code is a spin-$\frac{1}{2}$ Hamiltonian defined on a triangular lattice where the qubits are located at the vertices. The code consists of 3-qubit triangle operators acting at the vertices of the triangular plaquettes (Fig.~\ref{fig:model})
\begin{align}
\begin{split}
\hat{H}_{\textrm{XYZ}}& = \sum_{i=1}^N \left(g_1  \hat{T}_i^\nabla + g_2 \hat{T}_i^\Delta \right) \equiv \hat{H}_1 + \hat{H}_2,
\label{eq:H1H2Hamiltonian}
\end{split}
\end{align}
where $\hat{H}_1 \equiv \sum_{i=1}^N g_1  \hat{T}_i^\nabla$, $\hat{H}_2 \equiv \sum_{i=1}^N g_2 \hat{T}_i^\Delta$, and the triangle operators read $\hat{T}_k^\nabla \equiv \hat{Z}_k \hat{X}_i \hat{Y}_j$ and $T_j^\Delta \equiv \hat{Z}_j \hat{Y}_k \hat{X}_l$ where the orientation of the site indices denoted by the subscripts is shown in Fig.~\ref{fig:model}. We drop the hat notation in the rest of the paper, and consider the code in an $L \times L$ cluster with $n = L^2$ physical qubits.

Similarly to the Kitaev toric code~\cite{Kitaev2003}, $[H_1,H_2]=0$ and the Hamiltonian is a sum of two commuting parts, but the commutation properties are not as extensive as in the toric code. Instead, the Hamiltonian is strongly-frustrated because any up-up or down-down corner-sharing triangular plaquette operators anti-commute, which means that we are unable to simultaneously minimise the energy of every individual plaquette. The ground state is more complicated than a sum of individually satisfied plaquettes resembling more a quantum spin liquid~\cite{Majorana_Hubbard_Honeycomb}. We discuss elsewhere our numerical studies of the classical~\cite{TL} and quantum ground state~\cite{BTL}, emergent phenomena, symmetry group, and properties of the spectrum of the spin Hamiltonian $H_{\textrm{XYZ}}$~\cite{BTL}. 
Each of the two parts of the quantum Hamiltonian, $H_1$ and $H_2$, treated as a separate Hamiltonian possesses additionally a large number of non-Abelian plaquette-like symmetries which together with spatial symmetries we discuss elsewhere~\cite{BTL}. Here, we focus on the subsystem symmetries only.

\subsection{Generators of the stabiliser group $S$}

\begin{figure}[t]
        \centering
\begin{tikzpicture}[scale=0.3]
\definecolor{b1}{HTML}{5B7C99};
\definecolor{b2}{HTML}{FF8C42};
\definecolor{b3}{HTML}{CFE8EF};
\definecolor{b4}{HTML}{F4A261};

  \begin{scope}[y=(60:1),shift={(0,0)}]
          \foreach \x in {0,2}{
                    \draw[help lines,dashed,rotate=240]
    			  (\x+4,-4) -- (\x+4, -\x); 

          }
       		\foreach \x in {-4,-2,...,2}{
  				\draw[help lines,dashed]
    				(\x,-4) -- (\x,2) 
   					(-4,\x) -- (2,\x); 
          \draw[help lines,dashed,rotate=60]
    			  (\x+2,-2) -- (\x+2,-\x) ;
    				
			\foreach \y in {-8,-6,...,6}{
				\coordinate  (Dot\x\y) at (\x,\y);
    			}}
 			
              \draw [line width=0.4mm, draw=b2, opacity=1.0]
      (0,-2) -- (2,-2) node[draw,circle,inner sep=0.5pt,dotted,fill=white,line width=0.5mm]{$Z$} ;  

        \foreach \x in {-4,-2}{
        \draw [line width=0.4mm, draw=b2, opacity=1.0]
       			(\x,-2) node[draw,circle,inner sep=1pt,fill=white,line width=0.5mm]{$Z$}
       -- (\x+2,-2) node[draw,circle,inner sep=1pt,fill=white,line width=0.5mm]{$Z$};        
			}
			\draw (Dot-24)+(0,10mm) node[anchor=north]{$\Xi_i^Z$};    			
  \end{scope}
  
  \begin{scope}[y=(60:1),shift={(9,0)}]
          \foreach \x in {0,2}{
                    \draw[help lines,dashed,rotate=240]
    			  (\x+4,-4) -- (\x+4, -\x); 

          }
       		\foreach \x in {-4,-2,...,2}{
  				\draw[help lines,dashed]
    				(\x,-4) -- (\x,2) 
   					(-4,\x) -- (2,\x); 
          \draw[help lines,dashed,rotate=60]
    			  (\x+2,-2) -- (\x+2,-\x) ;
    				
			\foreach \y in {-8,-6,...,6}{
				\coordinate  (Dot\x\y) at (\x,\y);
    			}}
 			
              \draw [line width=0.4mm, draw=b2, opacity=1.0]
      (-2,0) -- (-2,2) node[draw,circle,inner sep=0.5pt,dotted,fill=white,line width=0.5mm]{$X$} ;  

        \foreach \x in {-4,-2}{
        \draw [line width=0.4mm, draw=b2, opacity=1.0]
       			(-2,\x) node[draw,circle,inner sep=1pt,fill=white,line width=0.5mm]{$X$}
       -- (-2,\x+2) node[draw,circle,inner sep=1pt,fill=white,line width=0.5mm]{$X$};        
			}
			\draw (Dot-24)+(0,10mm) node[anchor=north]{$\Xi_i^X$};    			
    			
  \end{scope}  

  \begin{scope}[y=(60:1),shift={(9+9,0)}]
          \foreach \x in {0,2}{
                    \draw[help lines,dashed,rotate=240]
    			  (\x+4,-4) -- (\x+4, -\x); 

          }
       		\foreach \x in {-4,-2,...,2}{
  				\draw[help lines,dashed]
    				(\x,-4) -- (\x,2) 
   					(-4,\x) -- (2,\x); 
          \draw[help lines,dashed,rotate=60]
    			  (\x+2,-2) -- (\x+2,-\x) ;
    				
			\foreach \y in {-8,-6,...,6}{
				\coordinate  (Dot\x\y) at (\x,\y);
    			}}
 			
              \draw [line width=0.4mm, draw=b2, opacity=1.0]
         (2,0) node[draw,circle,inner sep=0.5pt,dotted,fill=white,line width=0.5mm]{$Y$}              
      -- (0,2) node[draw,circle,inner sep=0.5pt,dotted,fill=white,line width=0.5mm]{$Y$} 
      (-6,2) -- (-4,0) node[draw,circle,inner sep=1pt,fill=white,line width=0.5mm]{$Y$}
			-- (-2,-2) node[draw,circle,inner sep=1pt,fill=white,line width=0.5mm]{$Y$}
			-- (0,-4) node[draw,circle,inner sep=1pt,fill=white,line width=0.5mm]{$Y$}

;
			\draw (Dot-24)+(0,10mm) node[anchor=north]{$\Xi_i^Y$};    			    			
  \end{scope}  
\end{tikzpicture}  
        \caption{\label{fig:LoopOperators}Single-loop operators $\Xi_i^Z$, $\Xi_i^X$, and $\Xi_i^Y$ that are conserved by the Majorana-XYZ Hamiltonian. The dashed site belongs to the next unit cell. The unit cell here is $3\times 3$.}              
    \end{figure}

We define here the stabiliser group $S$ of the Majorana-XYZ code, following the standard way to characterise a quantum error correction code by the stabiliser formalism~\cite{Gottesmann:Stabilizer,Poulin:Stabilizer}.

A direct calculation shows that the \textit{loop operators} $\Xi_l^{A}$ are symmetries, i.e. $\comm{H_{\textrm{XYZ}}}{\Xi_l^{A}}=0$, where $A \in \{X, Y, Z \}$ and the support is defined along a straight line in one of the three lattice directions as shown in Fig.~\ref{fig:LoopOperators}, looping around periodic boundary conditions. The index $l$ refers to the spatial position of the line. Note that a loop product of Pauli $A$ is a symmetry only in the direction shown in Fig.~\ref{fig:LoopOperators}. There are $3L$ loop operators, and they are similar to the symmetries along straight lines in the $90^\circ$ compass model~\cite{CompassModelsReview, Li:CompassCodes,nussinovDiscreteSlidingSymmetries2005}. In this context the code can be viewed as a generalisation to the triangular lattice of the quantum compass model on the square lattice.

In the thermodynamic limit or on tori which have no shear meaning the clusters considered here, each loop matches itself at the boundary. Loops of different kind $A \in \{X, Y, Z \}$ can intersect only once and therefore anticommute. Physically, the action of e.g. $\Xi_i^X$ flips one spin on the support of every $Z$-loop and $Y$-loop operator and thus flips the signs of their eigenvalues.

We define a \textit{double loop operator}, also a symmetry, as the product of two adjacent parallel loops $\Xi_i^A \Xi_{i+1}^A$. As double loop operators always intersect twice as often as one of its single loop constituents they satisfy
\begin{subequations}
    \label{eq:DLcommutation}
 \begin{align}
        \comm{\Xi_i^A \Xi_{i+1}^A}{\Xi_j^B \Xi_{j+1}^B} &=0\\ 
        \comm{\Xi_i^A \Xi_{i+1}^A}{\Xi_{j}^B} &=0
\end{align}   
\end{subequations}
for all line locations $i,j$ and all Pauli matrices $A\in \{X, Y, Z \}$ and $B\in \{X, Y, Z \}$.

The set of all double loops act as generators of the stabiliser group $S$,
\begin{equation}
    S = \left \langle \Xi_i^X \Xi_{i+1}^X,\Xi_i^Y \Xi_{i+1}^Y,\Xi_i^Z \Xi_{i+1}^Z \mid i \in \mathbb{Z}_{L-1} \right\rangle.
\end{equation}
The number of independent double-loop generators of $S$ is $|S|=3L - 3 - (L+1 \;\;\text{mod } 2)$ corresponding to an asymptotically constant encoding rate with $k = L^2 -|S|$ logical qubits in $L^2$ physical qubits. However, each stabiliser is of weight $2L$. For the subsystem variant the centre of the gauge group contains an additional stabiliser generator for even $L$ and the size of the centre is $|S|=3L - 3$ for all $L$.

\subsection{Identification as a subsystem code with $\frac{L}{2}$ logical qubits} 
Partitioning the Hilbert space as a subsystem code allows us to measure the stabilisers with only local measurements instead of weight $2L$ non-local measurements that span across the entire system, which comes at the cost of gauge degrees of freedom. In a subsystem code, the ambient Hilbert space $\mathcal{H}$ with $n=L^2$ physical qubits is partitioned as $\mathcal{H} = \left( \mathcal{L} \otimes \mathcal{G} \right) \otimes \mathcal{C}^\perp$, where the code space $\mathcal{C} = \mathcal{L} \otimes \mathcal{G}$ is divided into the logical space $\mathcal{L}$ containing logical qubits with the protected quantum information and the gauge space $\mathcal{G}$ containing gauge qubits. The gauge qubits are spin-$\frac{1}{2}$ degrees of freedom such that any two states related by a non-trivial transformation acting only on the gauge qubits are equivalent. The set of the gauge operators forms the gauge group $G$. 

For the Majorana-XYZ code, the local measurements to obtain the stabilisers are $3$-qubit measurements given by the triangle operators, which generate the gauge group $G$,
\begin{equation}
    G = \left \langle T_i^\nabla,T_j^\Delta \mid i,j \in \mathbb{Z}_{|G|_t} \right\rangle,
\end{equation}
where $|G|_t$ is the number of gauge triangles of type $t \in \{\nabla, \Delta\}$. 
Since any double loop is a product of the triangle operators, we can measure the double loops by measuring the triangle operators; even if the triangle operators do not commute, they are still constrained such that the product of the measurement results equals to the stabiliser state. The physical checks  correspond to the gauge operators and not to the stabiliser checks.

The centre of a group is the set of elements that commute with every element of the group. The centre of the gauge group $G$ is the stabiliser group $S$. The stabiliser group is the group of operators generated by the gauge operators that commute with every element of the gauge group, $S = \mathcal{Z}(G) \cap G$, where $\mathcal{Z}(G)$ is the centraliser of the gauge group $G$ in the Pauli group, i.e. the set of all Pauli operators that commute with every element of the group $G$. It can be instructive to re-write the gauge group $G$ as 
 \begin{equation}
 \begin{split} 
    G =& \left \langle  \Xi_j^X \Xi_{j+1}^X,\Xi_j^Y \Xi_{j+1}^Y,\Xi_j^Z \Xi_{j+1}^Z, X_{k}^{(g)}, Z_{k}^{(g)} \right. \\
    & \qquad \qquad \left. \mid j \in \mathbb{Z}_{L-1} , k \in \mathbb{Z}_{g} \right\rangle,
    \end{split}
\end{equation}
where $Z_k^{(g)}, X_k^{(g)}$ are the logical operators for the gauge qubits, given by products of the triangle operators that are not products of double loops (in general elements of the centre of $G$), and $g$ is the number of gauge qubits excluding stabilisers (see below).

Counting degrees of freedom, we have 
\begin{equation}
    k = n - g - |S|
\end{equation} 
logical qubits where $|S|$ is the size of the centre of $G$ i.e. the number of independent generators of stabiliser qubits, $g = (|G| - |S|)/2$ is the number of gauge qubits beyond the stabiliser qubits, and $n$ is the number of physical qubits, and $|A|$ is the cardinality, or the number of independent generators, of some group $A$. One possible choice for the independent generators of the gauge group $G$ 
are the Hamiltonian triangles independent of which there are $|G|=2L^2-3L+1$. We obtain $|G|$ numerically as the $F_2$ rank of $G$~\cite{open-source_xyz}. The centre of $G$ is of size $|S| = 3L-3$. It follows that we have $k=1$, i.e. a single logical qubit.

\subsection{Logical operators and topological encoding}

When the gauge group $G$ is non-Abelian, as it is here as not all Hamiltonian triangles commute, the `bare' logical operators are the set $L_\mathrm{b}= \mathcal{Z}(G) \symbol{92} S$, where the notation $A \symbol{92} B$ denotes the set of all the elements in $A$ that are not in $B$, and $\mathcal{Z}(G)$ is the centraliser of group $G$ in the Pauli group, i.e. the set of all Pauli operators that commute with every element of the group $G$. 

The `bare' logical operators are the single-line operators $\Xi_l^A$ as they commute with the Hamiltonian triangles as well as are excluded from the stabiliser group $S$ of double loops. Physically, it is impossible to get the single lines from triangle products, alluding to the fact that local errors cannot influence logical qubits. The logical operators always leave the stabiliser qubits intact, but in general need not leave the gauge qubits intact. Only the bare logical operators leave the gauge qubits intact. As a result of the gauge freedom of the subsystem code, we are free to act on the `bare' logical operators by any gauge operator giving `dressed' logical operators.  The set of all such dressed logical operators is given by the subset of the centralizer of the stabiliser subgroup that excludes all elements in the gauge group, $L(G) = \mathcal{Z}(S) \symbol{92} G $ i.e. all Pauli operators that commute with all the double loop stabilisers and their products but are not the Hamiltonian triangles or their products (these products include the double loops but not single loops). The distance $d$ is the weight of the smallest operator in $L(G)$; for the Majorana-XYZ code $d=L$ given by the weight of a single-line $\Xi_l^A$, which corresponds to the smallest length of a homologically non-trivial cycle.

The code space is a specific symmetry sector where the defining symmetries are the double loops. No local measurement can transform topological ground states into each other: local gauge perturbations can only increase the length of the line operators but not change the winding. Indeed, in topological order different degenerate ground states cannot be distinguished by a local measurement. Here, we see quite directly how exotic states of matter like the topologically ordered system forming the code space can be used to robustly store and process quantum information.

\subsection{Threshold estimation}
\begin{figure}[t]
\includegraphics[scale=0.55]{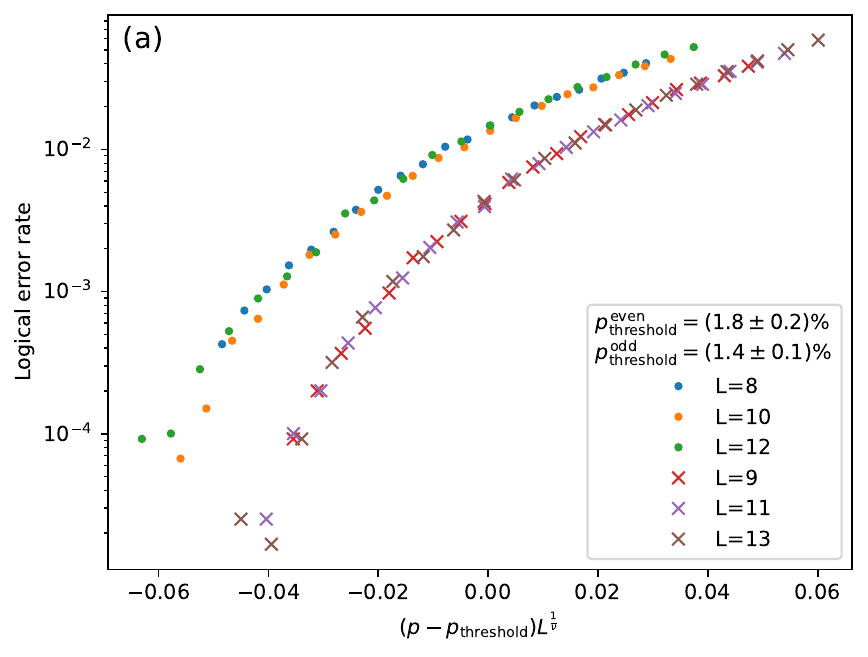}
\includegraphics[scale=0.25]{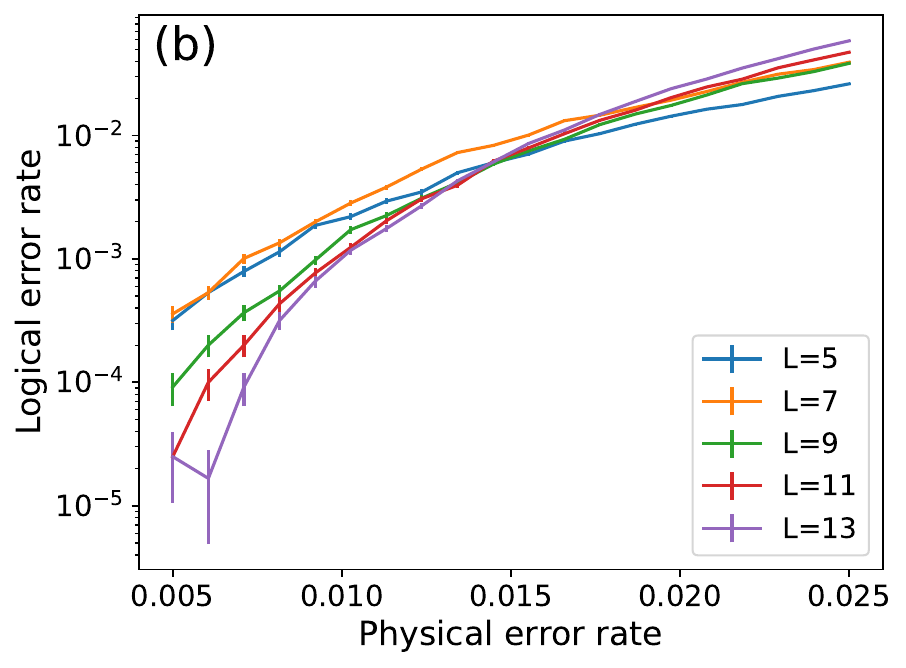}
\includegraphics[scale=0.25]{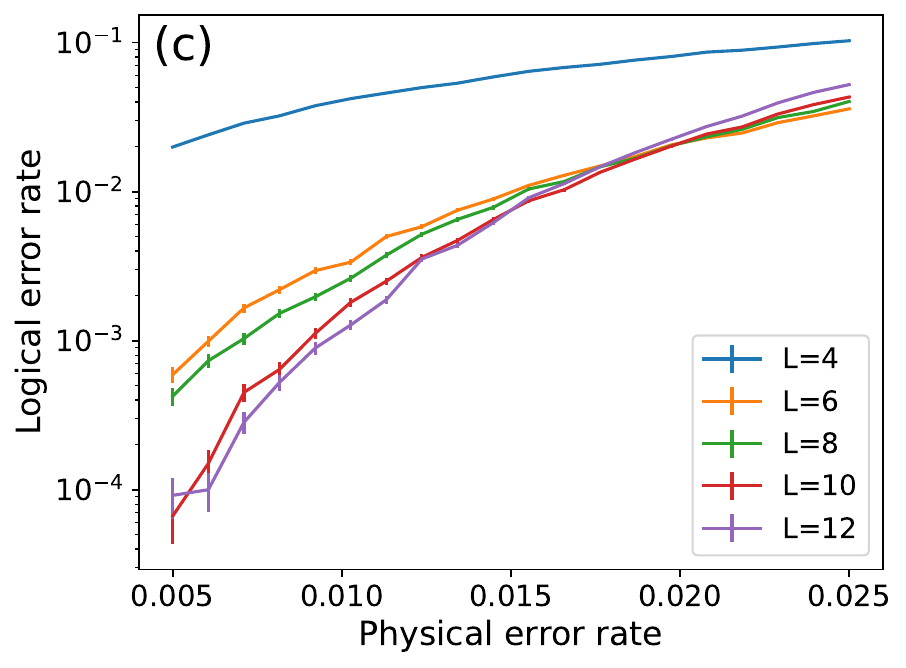}
\caption{ \label{fig:threshold} Threshold estimation. (a) Finite-size scaling results in $p_\mathrm{threshold}^\mathrm{odd} = 1.4\% \pm 0.1\%$ for odd distance and $p_\mathrm{threshold}^\mathrm{even} = 1.8\% \pm 0.2\%$ for even distance. Here the quality of fits are $\chi^2_\mathrm{even} = 0.71$ and $\chi^2_\mathrm{odd} = 4.06$ with $\chi^2=1$ indicating a perfect fit with the assumed uncertainties. We solve the optimisation problem using the python package `fssa', which uses the reduced $\chi^2$ quality function of Ref.~\cite{PhysRevB.70.014418} to measure the quality of the data collapse. The numerically fitted critical exponents are $\nu_\mathrm{even} = 1.5 \pm 1.0$ and $\nu_\mathrm{odd} = 1.5\pm 0.5$. (b) Unscaled logical error vs. physical error for odd distances displaying finite-size effects. (c) Same as (b) but for even distances.}
\end{figure}

We consider independent $X,Z$ noise that occurs at probability $p$ per qubit. The resulting Tanner graph has degree 6, for decoding of which we use Monte Carlo sampling using the belief propagation with ordered statistics decoder (BP-OSD)~\cite{roffe_decoding_2020,Roffe_LDPC_Python_tools_2022}. We use 120000 shots per $p, d$, and consider large-distance codes upto $d = 13$~\cite{open-source_xyz}. We observe finite-size scaling (Fig.~\ref{fig:threshold}) meaning small distances overestimate the threshold. The apparent threshold error rate decreases with increasing distance because the larger-$L$ lines intersect at lower physical error rates. To obtain the threshold we collapse the data using 
\begin{equation}
p_\mathrm{logical} = f(p - p_\mathrm{threshold}) L^{\frac{1}{\nu}},
\end{equation}
which holds in the infinite-distance limit $L\to\infty$ and $p\to p_\mathrm{threshold}$, where $f$ is an unknown scaling function, $\nu$ is an unknown critical exponent, and $ p_\mathrm{threshold}$ is an unknown scaling parameter corresponding to the phase transition~\cite{PhysRevA.81.032301,Dennis2002,Wang2003} separating the two regimes where errors proliferate and where stabiliser order is maintained.

The observed logical error rates differ between odd and even distance (Fig.~\ref{fig:threshold} (b) and (c)). One possible reason is that for odd distance the stabilisers are all double loops, but for even distance one double loop is linearly dependent and there is one additional non-double-loop generator. For the even case we provide the decoder with an overcomplete stabiliser check matrix where the number of stabiliser generators is one higher than the rank over $F_2$. We speculate that using dependent checks improves the decoder's convergence reducing ambiguous syndromes that the BP-OSD decoder miscorrects into a logical operation by adding extra constraints that help break degeneracies and avoid convergence traps.

We leave a detailed study of different decoding mechanisms, noise profiles including circuit-level sampling, and considering the effect on the threshold of the order in which to perform the gauge triangle measurements, to later work. For example, a modification of the surface code under biased $Z$-noise has been observed to result in high thresholds~\cite{PhysRevLett.120.050505}. Here, notably, we do not have the same locality constraints for stabilisers as in traditional topological codes~\cite{PhysRevLett.120.050505}. Another important scenario is a circuit model where the native operations are three-qubit measurements similar to the two-qubit native measurements in the honeycomb code studied in Ref.~\cite{Gidney2021faulttolerant}, implemented by e.g. slightly longer measurement loops than the two-qubit operations in the Majorana chip~\cite{2025MSFT_Majorana-1,PRXQuantum.4.010310}. 

\section{Outlook and further work}

\begin{table}[t]
\caption{Code parameters. Number of physical qubits is $n=L^2$.}
\label{tab:counts}
\begin{ruledtabular}
\begin{tabular}{cccc}
Code & Logical qubits $k$ & Distance $d$ & Threshold  \\
\hline
Non-subsystem & $n-O(\sqrt{n})$  & 3 & No scalable $d$  \\
Subsystem & 1 & $\sqrt{n}$ & 
\begin{tabular}[t]{@{}c@{}}
$(1.8\pm 0.2)\% \, (\text{even } d)$\\
$(1.4\pm 0.1)\% \, (\text{odd } d)$
\end{tabular}
 \\
\end{tabular}
\end{ruledtabular}
\end{table}

We present a quantum error correction surface subsystem code that is most native to Majorana-based qubit modalities where the native operations are few-qubit measurements. The code is geometrically implemented on qubits in a triangular lattice with local 3-qubit interaction terms, and the single-qubit logical gates, represented by homologically non-trivial cycles, are transversal (Fig.~\ref{fig:LoopOperators}). The threshold of around $1.6 \%$ compares favourably with modern surface codes and in particular with subsystem codes. If the triangle operators are not treated as gauge generators we obtain a constant distance $d=3$ and a constant encoding rate $k/n \to 1$ as $L\to\infty$ with $k=n - |S|$ logical qubits (Table~\ref{tab:counts}).

It is easy to show that the magnetisation and all connected 2-point correlators vanish (Appendix) meaning the code can detect and correct all 1-qubit and all 2-qubit errors (as long as $d>4$). Higher-order correlated errors of weight equal to or below $\lfloor (d-1)/2\rfloor \sim O(\sqrt{n})$ are all correctable assuming they are not in the gauge group, which however leaves the logical qubit intact. Errors that cannot be corrected are also unlikely to occur due to being larger than $O(\sqrt{n})$ in weight, similarly as in topological codes~\cite{https://doi.org/10.48550/arxiv.1311.0277}, reflecting the below-threshold behaviour where fault-tolerance can be achieved with asymptotically perfect accuracy in the thermodynamic limit (Fig.~\ref{fig:threshold} (b) and (c)).
 
Open boundary conditions can be a better match with hardware and not necessarily deteriorate the code performance~\cite{Gidney2022benchmarkingplanar}. For the Majorana-XYZ code this involves cutting the triangular lattice into a large macroscopic triangle with three distinct boundaries along which we must allow the bulk logical string operators to terminate safely. The gauge triangles at the boundaries must be modified, and to prevent frustration at the corners, we need to measure operators that accommodate both edges. We leave for later work the study of the bulk string excitations in an open triangular patch, and the topological bulk-boundary dynamics including the details how the boundary triangles must be measured, e.g. considering two-qubit boundary $ZZ, YY, XX$ terms along the corresponding edge types as in Fig.~\ref{fig:LoopOperators}. 
By restricting to fixed-size triangular patches of the Majorana-XYZ code with a single logical qubit in each patch, measuring longer single lines across patches forms the basis of lattice surgery~\cite{Horsman2012,Chatterjee2025}, for e.g. the implementation of the CNOT gate. We leave for later study the logical non-Clifford gates and a universal logical gate set, including code switching~\cite{PhysRevLett.113.080501} between subsystem codes in different gauges.

We leave for later work the generalisation of the code to a Floquet subsystem variant~\cite{Hastings2021} with dynamical logical qubits. The gauge triangle plaquettes can be divided into three mutually anti-commuting sublattices $A$, $B$, and $C$, where all triangles in a given sublattice commute. Measuring them in a periodic sequence $A \rightarrow B \rightarrow C \rightarrow A$ defines a 3-step Floquet code cycle. Only the homologically non-trivial operators survive the full cycle. The additional logical qubits exist in the space freed by creating a gauge defect, which arises from the non-commuting check measurements here represented by a sequence of gauge-fixing measurements of the gauge triangles which don't mutually all commute.

\begin{acknowledgments}
L.T. is supported by the Österreichischer Wissenschaftsfonds FWF (M 2939). We would like to thank Andreas L\"auchli for fruitful discussions.

\end{acknowledgments}

\appendix

\section{Connection to the Majorana-Hubbard model on the honeycomb lattice}
The Majorana-XYZ code can be equivalently expressed in terms of Majorana fermions on the honeycomb lattice. This system was originally considered by Li and Franz~\cite{Majorana_Hubbard_Honeycomb}. The honeycomb lattice has two sites per unit cell, and we label the Majorana operators by $\alpha_i$ and $\beta_j$ associated with the lattice sites $i$ and $j$ of the two sublattices of a honeycomb lattice respectively. Essentially, we can imagine the site index labeling vertical links (Fig.~\ref{fig:model}). 

Taking nearest-neighbour Majorana interactions on the honeycomb lattice, the simplest symmetry allowed Hamiltonian has 4-Majorana interaction terms, reading $H = H_0 + H_{\textrm{int}}$, where 
\begin{align}
\begin{split}
H_0 &= \rmi t \sum_{\langle i j \rangle} \eta_{ij} \alpha_i \beta_j \hspace{.5cm} \hspace{1cm} \textrm{and} \\
H_{\textrm{int}} &= g_1 \sum_{\ysym{scale=0.1}} \alpha_i \beta_j \beta_k \beta_i + g_2 \sum_{\lambdasym{scale=0.1}} \beta_j \alpha_j \alpha_k  \alpha_l,
\label{eq:MajoranaHubbardHamiltonian}
\end{split}
\end{align}
where $t, g_1, g_2$ are real coupling constants.
The sum in the hopping Hamiltonian $H_0$ is defined over adjacent sites $\langle i j \rangle$ and the phase factors $\eta_{ij}$ are constrained by the Grosfeld-Stern rule. We can set $\eta_{ij} = 1$ using a suitable choice for hopping orientations~\cite{Majorana_Hubbard_Honeycomb}.
The interacting terms in $H_{\textrm{int}}$ are defined to be four-point Majorana interactions arranged along a $\ysym{scale=0.1}$- and a $\lambdasym{scale=0.1}$-shape (Fig.~\ref{fig:model}). The Majorana-XYZ corresponds to the $t = 0$ strong interacting limit, possible to achieve experimentally by adjusting e.g. the Majorana-Majorana separation~\cite{toikka_2019}.

To see this, we choose a dimerization pattern on the honeycomb lattice where two Majorana sites are combined into a single spinless fermion site. If the dimerization pattern is a particular staggered dimer state on the honeycomb lattice, then spinless fermions reside on a triangular lattice. In the thermodynamic limit, a Jordan-Wigner transformation can be used to map the spinless fermions and their interactions into the (bosonic) Majorana-XYZ Hamiltonian with three-spin interactions~\cite{Majorana_Hubbard_Honeycomb}. In finite-size clusters with periodic boundary conditions some care needs to be taken in order to faithfully map the spectrum of the Majorana-Hubbard model to an equivalent spin model. Here, for simplicity, we ignore this effect and study the three-spin Majorana-XYZ Hamiltonian with periodic boundary conditions. For completeness, assuming $k<i<j$ and $i+1=j$, the $\ysym{scale=0.1}$-shaped term $g_1 \alpha_k \beta_i \beta_j$ is mapped to the local Pauli operator $T_k^\nabla \equiv Z_k X_i Y_j$. Analogously, the $\lambdasym{scale=0.1}$-shaped term $g_2 \beta_i \alpha_j \alpha_k \alpha_l$ is mapped onto the triangular operator $T_j^\Delta \equiv Z_j Y_k X_l$ when $k<l<j$ and $k+1=l$.

\section{Detectable and undetectable errors}
The dressing corresponds to an undetectable error. Action by topologically non-trivial elements in the gauge group transforms dressed logical operators into each other, and therefore cannot be detected as errors by the code. Any dressed logical operator automatically commutes with all double loops. Of course, any dressed operator is an equally valid choice.
 
From a condensed matter perspective, the Majorana-XYZ code defined in Eq.~\eqref{eq:H1H2Hamiltonian} has a rich symmetry group. Importantly, the symmetry structure immediately results in many vanishing correlation functions, which have a physical meaning within the code as detectable physical errors. Non-vanishing expectations correspond to undetectable errors. Detectable errors live in the space $P_n \symbol{92} \mathcal{Z}(S)$ where $P_n$ is the Pauli group

The common $+1$ eigenspace of the stabilisers defines the codespace $\mathcal{C}$ of the error correction code and any operator whose expectation value vanishes corresponds to a detectable error. A detectable error anticommutes with at least one stabiliser. The codespace is a specifically chosen symmetry sector and the correlation functions may only be non-zero if the operator in consideration acts non-trivially within the subspace. Products of the gauge operators that do not lie in $S$ will have non-vanishing expectation values.

\subsection{Errors of weight 1 (magnetization)}
Let $A_j$ be a Pauli matrix $X$, $Y$ or $Z$ acting on a site $j$. 
On any lattice there is a loop operator  $\Xi_i^{B}$ with $B \neq A$ and $i$ chosen such that the loop runs over the site $j$. Then, $\acomm{A_j}{\Xi_i^{B}}=0$ and it follows that $\langle A_j \rangle = 0$ for all Pauli matrices $A \in \{X, Y, Z \}$. The spins do not have a preferred direction and the magnetization vanishes.

\subsection{Errors of weight 2}

Let $A_i B_j$ be a Pauli operator acting non-trivially on two sites $i$ and $j$. Then the two Pauli matrices $A$ and $B$ may be either the same or different. It is always possible to find a loop operator $\Xi_k^A$ that anti-commutes with one Pauli but commutes with the other, except when $i = j$ and $A = B$. All connected 2-point correlators vanish, meaning any pair of spins is uncorrelated.

All errors of weight 1 and 2 are therefore detectable and correctable by the Majorana-XYZ code.

\subsection{Errors of weight 3 and higher }
It is always possible to find a loop operator $\Xi_k^A$ such that it anti-commutes with an odd number of Paulis but commutes with the rest, except when the error is a triangle operator or their scale up. The scale up means a larger 3-qubit triangle congruent to $T^\nabla$ or $T^\Delta$.

\subsection{Higher correlations}
\label{subsec:spincorrx}
\begin{figure}[htpb]
\includegraphics[scale=1]{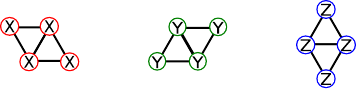}
\caption{ \label{fig:weight4} Three undetectable errors of weight four. All such errors belong to the gauge group (products of triangle operators) and commute with all stabilisers.}
\end{figure}

The four-point correlators that do not vanish correspond to operators that are products of two edge-sharing triangle operators: To commute with the loop operators, we need an even number of Pauli operators of the opposite types along the X-, Y- and Z- directions. Here we find diamond and parallelogram shaped operators, which, again are products of the triangle operators (Fig.~\ref{fig:weight4}).

\bibliographystyle{apsrev4-1}
\bibliography{xyz_bib}

\begin{thebibliography}{46}%
\makeatletter
\providecommand \@ifxundefined [1]{%
 \@ifx{#1\undefined}
}%
\providecommand \@ifnum [1]{%
 \ifnum #1\expandafter \@firstoftwo
 \else \expandafter \@secondoftwo
 \fi
}%
\providecommand \@ifx [1]{%
 \ifx #1\expandafter \@firstoftwo
 \else \expandafter \@secondoftwo
 \fi
}%
\providecommand \natexlab [1]{#1}%
\providecommand \enquote  [1]{``#1''}%
\providecommand \bibnamefont  [1]{#1}%
\providecommand \bibfnamefont [1]{#1}%
\providecommand \citenamefont [1]{#1}%
\providecommand \href@noop [0]{\@secondoftwo}%
\providecommand \href [0]{\begingroup \@sanitize@url \@href}%
\providecommand \@href[1]{\@@startlink{#1}\@@href}%
\providecommand \@@href[1]{\endgroup#1\@@endlink}%
\providecommand \@sanitize@url [0]{\catcode `\\12\catcode `\$12\catcode
  `\&12\catcode `\#12\catcode `\^12\catcode `\_12\catcode `\%12\relax}%
\providecommand \@@startlink[1]{}%
\providecommand \@@endlink[0]{}%
\providecommand \url  [0]{\begingroup\@sanitize@url \@url }%
\providecommand \@url [1]{\endgroup\@href {#1}{\urlprefix }}%
\providecommand \urlprefix  [0]{URL }%
\providecommand \Eprint [0]{\href }%
\providecommand \doibase [0]{http://dx.doi.org/}%
\providecommand \selectlanguage [0]{\@gobble}%
\providecommand \bibinfo  [0]{\@secondoftwo}%
\providecommand \bibfield  [0]{\@secondoftwo}%
\providecommand \translation [1]{[#1]}%
\providecommand \BibitemOpen [0]{}%
\providecommand \bibitemStop [0]{}%
\providecommand \bibitemNoStop [0]{.\EOS\space}%
\providecommand \EOS [0]{\spacefactor3000\relax}%
\providecommand \BibitemShut  [1]{\csname bibitem#1\endcsname}%
\let\auto@bib@innerbib\@empty
\bibitem [{\citenamefont {Breuckmann}\ and\ \citenamefont
  {Eberhardt}(2021)}]{PRXQuantum.2.040101}%
  \BibitemOpen
  \bibfield  {author} {\bibinfo {author} {\bibfnamefont {N.~P.}\ \bibnamefont
  {Breuckmann}}\ and\ \bibinfo {author} {\bibfnamefont {J.~N.}\ \bibnamefont
  {Eberhardt}},\ }\href {\doibase 10.1103/PRXQuantum.2.040101} {\bibfield
  {journal} {\bibinfo  {journal} {PRX Quantum}\ }\textbf {\bibinfo {volume}
  {2}},\ \bibinfo {pages} {040101} (\bibinfo {year} {2021})}\BibitemShut
  {NoStop}%
\bibitem [{\citenamefont {Gottesman}(2014)}]{Gottesman2014}%
  \BibitemOpen
  \bibfield  {author} {\bibinfo {author} {\bibfnamefont {D.}~\bibnamefont
  {Gottesman}},\ }\href {\doibase 10.26421/qic14.15-16-5} {\bibfield  {journal}
  {\bibinfo  {journal} {Quantum Inf. Comput.}\ }\textbf {\bibinfo {volume}
  {14}},\ \bibinfo {pages} {1339} (\bibinfo {year} {2014})}\BibitemShut
  {NoStop}%
\bibitem [{\citenamefont {Knill}(2005)}]{Knill2005}%
  \BibitemOpen
  \bibfield  {author} {\bibinfo {author} {\bibfnamefont {E.}~\bibnamefont
  {Knill}},\ }\href {\doibase 10.1038/nature03350} {\bibfield  {journal}
  {\bibinfo  {journal} {Nature}\ }\textbf {\bibinfo {volume} {434}},\ \bibinfo
  {pages} {39–44} (\bibinfo {year} {2005})}\BibitemShut {NoStop}%
\bibitem [{\citenamefont {Fowler}\ \emph {et~al.}(2012)\citenamefont {Fowler},
  \citenamefont {Mariantoni}, \citenamefont {Martinis},\ and\ \citenamefont
  {Cleland}}]{PhysRevA.86.032324}%
  \BibitemOpen
  \bibfield  {author} {\bibinfo {author} {\bibfnamefont {A.~G.}\ \bibnamefont
  {Fowler}}, \bibinfo {author} {\bibfnamefont {M.}~\bibnamefont {Mariantoni}},
  \bibinfo {author} {\bibfnamefont {J.~M.}\ \bibnamefont {Martinis}}, \ and\
  \bibinfo {author} {\bibfnamefont {A.~N.}\ \bibnamefont {Cleland}},\ }\href
  {\doibase 10.1103/PhysRevA.86.032324} {\bibfield  {journal} {\bibinfo
  {journal} {Phys. Rev. A}\ }\textbf {\bibinfo {volume} {86}},\ \bibinfo
  {pages} {032324} (\bibinfo {year} {2012})}\BibitemShut {NoStop}%
\bibitem [{\citenamefont {Wang}\ \emph {et~al.}(2011)\citenamefont {Wang},
  \citenamefont {Fowler},\ and\ \citenamefont
  {Hollenberg}}]{PhysRevA.83.020302}%
  \BibitemOpen
  \bibfield  {author} {\bibinfo {author} {\bibfnamefont {D.~S.}\ \bibnamefont
  {Wang}}, \bibinfo {author} {\bibfnamefont {A.~G.}\ \bibnamefont {Fowler}}, \
  and\ \bibinfo {author} {\bibfnamefont {L.~C.~L.}\ \bibnamefont
  {Hollenberg}},\ }\href {\doibase 10.1103/PhysRevA.83.020302} {\bibfield
  {journal} {\bibinfo  {journal} {Phys. Rev. A}\ }\textbf {\bibinfo {volume}
  {83}},\ \bibinfo {pages} {020302(R)} (\bibinfo {year} {2011})}\BibitemShut
  {NoStop}%
\bibitem [{\citenamefont {Raussendorf}\ \emph {et~al.}(2007)\citenamefont
  {Raussendorf}, \citenamefont {Harrington},\ and\ \citenamefont
  {Goyal}}]{Raussendorf_2007}%
  \BibitemOpen
  \bibfield  {author} {\bibinfo {author} {\bibfnamefont {R.}~\bibnamefont
  {Raussendorf}}, \bibinfo {author} {\bibfnamefont {J.}~\bibnamefont
  {Harrington}}, \ and\ \bibinfo {author} {\bibfnamefont {K.}~\bibnamefont
  {Goyal}},\ }\href {\doibase 10.1088/1367-2630/9/6/199} {\bibfield  {journal}
  {\bibinfo  {journal} {New J. Phys.}\ }\textbf {\bibinfo {volume} {9}},\
  \bibinfo {pages} {199} (\bibinfo {year} {2007})}\BibitemShut {NoStop}%
\bibitem [{\citenamefont {Aliferis}\ \emph {et~al.}(2005)\citenamefont
  {Aliferis}, \citenamefont {Gottesman},\ and\ \citenamefont
  {Preskill}}]{aliferis2005quantumaccuracythresholdconcatenated}%
  \BibitemOpen
  \bibfield  {author} {\bibinfo {author} {\bibfnamefont {P.}~\bibnamefont
  {Aliferis}}, \bibinfo {author} {\bibfnamefont {D.}~\bibnamefont {Gottesman}},
  \ and\ \bibinfo {author} {\bibfnamefont {J.}~\bibnamefont {Preskill}},\
  }\href {https://arxiv.org/abs/quant-ph/0504218} {} (\bibinfo {year} {2005}),\
  \Eprint {http://arxiv.org/abs/quant-ph/0504218} {arXiv:quant-ph/0504218
  [quant-ph]} \BibitemShut {NoStop}%
\bibitem [{\citenamefont {Gidney}\ \emph {et~al.}(2021)\citenamefont {Gidney},
  \citenamefont {Newman}, \citenamefont {Fowler},\ and\ \citenamefont
  {Broughton}}]{Gidney2021faulttolerant}%
  \BibitemOpen
  \bibfield  {author} {\bibinfo {author} {\bibfnamefont {C.}~\bibnamefont
  {Gidney}}, \bibinfo {author} {\bibfnamefont {M.}~\bibnamefont {Newman}},
  \bibinfo {author} {\bibfnamefont {A.}~\bibnamefont {Fowler}}, \ and\ \bibinfo
  {author} {\bibfnamefont {M.}~\bibnamefont {Broughton}},\ }\href {\doibase
  10.22331/q-2021-12-20-605} {\bibfield  {journal} {\bibinfo  {journal}
  {{Quantum}}\ }\textbf {\bibinfo {volume} {5}},\ \bibinfo {pages} {605}
  (\bibinfo {year} {2021})}\BibitemShut {NoStop}%
\bibitem [{\citenamefont {Higgott}\ and\ \citenamefont
  {Breuckmann}(2021)}]{PhysRevX.11.031039}%
  \BibitemOpen
  \bibfield  {author} {\bibinfo {author} {\bibfnamefont {O.}~\bibnamefont
  {Higgott}}\ and\ \bibinfo {author} {\bibfnamefont {N.~P.}\ \bibnamefont
  {Breuckmann}},\ }\href {\doibase 10.1103/PhysRevX.11.031039} {\bibfield
  {journal} {\bibinfo  {journal} {Phys. Rev. X}\ }\textbf {\bibinfo {volume}
  {11}},\ \bibinfo {pages} {031039} (\bibinfo {year} {2021})}\BibitemShut
  {NoStop}%
\bibitem [{\citenamefont {Paetznick}\ \emph {et~al.}(2023)\citenamefont
  {Paetznick}, \citenamefont {Knapp}, \citenamefont {Delfosse}, \citenamefont
  {Bauer}, \citenamefont {Haah}, \citenamefont {Hastings},\ and\ \citenamefont
  {da~Silva}}]{PRXQuantum.4.010310}%
  \BibitemOpen
  \bibfield  {author} {\bibinfo {author} {\bibfnamefont {A.}~\bibnamefont
  {Paetznick}}, \bibinfo {author} {\bibfnamefont {C.}~\bibnamefont {Knapp}},
  \bibinfo {author} {\bibfnamefont {N.}~\bibnamefont {Delfosse}}, \bibinfo
  {author} {\bibfnamefont {B.}~\bibnamefont {Bauer}}, \bibinfo {author}
  {\bibfnamefont {J.}~\bibnamefont {Haah}}, \bibinfo {author} {\bibfnamefont
  {M.~B.}\ \bibnamefont {Hastings}}, \ and\ \bibinfo {author} {\bibfnamefont
  {M.~P.}\ \bibnamefont {da~Silva}},\ }\href {\doibase
  10.1103/PRXQuantum.4.010310} {\bibfield  {journal} {\bibinfo  {journal} {PRX
  Quantum}\ }\textbf {\bibinfo {volume} {4}},\ \bibinfo {pages} {010310}
  (\bibinfo {year} {2023})}\BibitemShut {NoStop}%
\bibitem [{\citenamefont {Brown}\ \emph {et~al.}(2016)\citenamefont {Brown},
  \citenamefont {Nickerson},\ and\ \citenamefont {Browne}}]{Brown2016-dq}%
  \BibitemOpen
  \bibfield  {author} {\bibinfo {author} {\bibfnamefont {B.~J.}\ \bibnamefont
  {Brown}}, \bibinfo {author} {\bibfnamefont {N.~H.}\ \bibnamefont
  {Nickerson}}, \ and\ \bibinfo {author} {\bibfnamefont {D.~E.}\ \bibnamefont
  {Browne}},\ }\href {\doibase https://doi.org/10.1038/ncomms12302} {\bibfield
  {journal} {\bibinfo  {journal} {Nat. Commun.}\ }\textbf {\bibinfo {volume}
  {7}},\ \bibinfo {pages} {12302} (\bibinfo {year} {2016})}\BibitemShut
  {NoStop}%
\bibitem [{\citenamefont {Wang}\ \emph
  {et~al.}(2003{\natexlab{a}})\citenamefont {Wang}, \citenamefont
  {Harrington},\ and\ \citenamefont {Preskill}}]{Wang2003-zy}%
  \BibitemOpen
  \bibfield  {author} {\bibinfo {author} {\bibfnamefont {C.}~\bibnamefont
  {Wang}}, \bibinfo {author} {\bibfnamefont {J.}~\bibnamefont {Harrington}}, \
  and\ \bibinfo {author} {\bibfnamefont {J.}~\bibnamefont {Preskill}},\ }\href
  {\doibase https://doi.org/10.1016/S0003-4916(02)00019-2} {\bibfield
  {journal} {\bibinfo  {journal} {Ann. Phys. (N. Y.)}\ }\textbf {\bibinfo
  {volume} {303}},\ \bibinfo {pages} {31} (\bibinfo {year}
  {2003}{\natexlab{a}})}\BibitemShut {NoStop}%
\bibitem [{\citenamefont {Kitaev}(2006)}]{Kitaev2006}%
  \BibitemOpen
  \bibfield  {author} {\bibinfo {author} {\bibfnamefont {A.}~\bibnamefont
  {Kitaev}},\ }\href {\doibase 10.1016/j.aop.2005.10.005} {\bibfield  {journal}
  {\bibinfo  {journal} {Ann. Phys.}\ }\textbf {\bibinfo {volume} {321}},\
  \bibinfo {pages} {2–111} (\bibinfo {year} {2006})}\BibitemShut {NoStop}%
\bibitem [{\citenamefont {Lee}\ \emph {et~al.}(2017)\citenamefont {Lee},
  \citenamefont {Brell},\ and\ \citenamefont {Flammia}}]{Lee_2017}%
  \BibitemOpen
  \bibfield  {author} {\bibinfo {author} {\bibfnamefont {Y.-C.}\ \bibnamefont
  {Lee}}, \bibinfo {author} {\bibfnamefont {C.~G.}\ \bibnamefont {Brell}}, \
  and\ \bibinfo {author} {\bibfnamefont {S.~T.}\ \bibnamefont {Flammia}},\
  }\href {\doibase 10.1088/1742-5468/aa7ee2} {\bibfield  {journal} {\bibinfo
  {journal} {Journal of Statistical Mechanics: Theory and Experiment}\ }\textbf
  {\bibinfo {volume} {2017}},\ \bibinfo {pages} {083106} (\bibinfo {year}
  {2017})}\BibitemShut {NoStop}%
\bibitem [{\citenamefont {Hastings}\ and\ \citenamefont
  {Haah}(2021)}]{Hastings2021}%
  \BibitemOpen
  \bibfield  {author} {\bibinfo {author} {\bibfnamefont {M.~B.}\ \bibnamefont
  {Hastings}}\ and\ \bibinfo {author} {\bibfnamefont {J.}~\bibnamefont
  {Haah}},\ }\href {\doibase 10.22331/q-2021-10-19-564} {\bibfield  {journal}
  {\bibinfo  {journal} {Quantum}\ }\textbf {\bibinfo {volume} {5}},\ \bibinfo
  {pages} {564} (\bibinfo {year} {2021})}\BibitemShut {NoStop}%
\bibitem [{\citenamefont {Aghaee}\ \emph {et~al.}(2025)\citenamefont {Aghaee}
  \emph {et~al.}}]{2025MSFT_Majorana-1}%
  \BibitemOpen
  \bibfield  {author} {\bibinfo {author} {\bibfnamefont {M.}~\bibnamefont
  {Aghaee}} \emph {et~al.},\ }\href {\doibase 10.1038/s41586-024-08445-2}
  {\bibfield  {journal} {\bibinfo  {journal} {Nature}\ }\textbf {\bibinfo
  {volume} {638}},\ \bibinfo {pages} {651–655} (\bibinfo {year}
  {2025})}\BibitemShut {NoStop}%
\bibitem [{\citenamefont {Toikka}(2019)}]{toikka_2019}%
  \BibitemOpen
  \bibfield  {author} {\bibinfo {author} {\bibfnamefont {L.~A.}\ \bibnamefont
  {Toikka}},\ }\href {\doibase 10.1088/1367-2630/ab5336} {\bibfield  {journal}
  {\bibinfo  {journal} {New J. Phys.}\ }\textbf {\bibinfo {volume} {21}},\
  \bibinfo {pages} {113033} (\bibinfo {year} {2019})}\BibitemShut {NoStop}%
\bibitem [{\citenamefont {Li}\ and\ \citenamefont
  {Franz}(2018)}]{Majorana_Hubbard_Honeycomb}%
  \BibitemOpen
  \bibfield  {author} {\bibinfo {author} {\bibfnamefont {C.}~\bibnamefont
  {Li}}\ and\ \bibinfo {author} {\bibfnamefont {M.}~\bibnamefont {Franz}},\
  }\href {\doibase 10.1103/PhysRevB.98.115123} {\bibfield  {journal} {\bibinfo
  {journal} {Physical Review B}\ }\textbf {\bibinfo {volume} {98}},\ \bibinfo
  {pages} {115123} (\bibinfo {year} {2018})}\BibitemShut {NoStop}%
\bibitem [{\citenamefont {Busse}\ \emph {et~al.}()\citenamefont {Busse},
  \citenamefont {Toikka},\ and\ \citenamefont {L\"{a}uchli}}]{BTL}%
  \BibitemOpen
  \bibfield  {author} {\bibinfo {author} {\bibfnamefont {T.}~\bibnamefont
  {Busse}}, \bibinfo {author} {\bibfnamefont {L.}~\bibnamefont {Toikka}}, \
  and\ \bibinfo {author} {\bibfnamefont {A.}~\bibnamefont {L\"{a}uchli}},\
  }\href@noop {} {\enquote {\bibinfo {title} {Unpublished},}\ }\BibitemShut
  {NoStop}%
\bibitem [{\citenamefont {Zhou}\ \emph {et~al.}(2011)\citenamefont {Zhou},
  \citenamefont {Zhang},\ and\ \citenamefont {Yi}}]{zhou2011}%
  \BibitemOpen
  \bibfield  {author} {\bibinfo {author} {\bibfnamefont {J.}~\bibnamefont
  {Zhou}}, \bibinfo {author} {\bibfnamefont {W.}~\bibnamefont {Zhang}}, \ and\
  \bibinfo {author} {\bibfnamefont {W.}~\bibnamefont {Yi}},\ }\href {\doibase
  10.1103/PhysRevA.84.063603} {\bibfield  {journal} {\bibinfo  {journal} {Phys.
  Rev. A}\ }\textbf {\bibinfo {volume} {84}},\ \bibinfo {pages} {063603}
  (\bibinfo {year} {2011})}\BibitemShut {NoStop}%
\bibitem [{\citenamefont {Volovik}(1999)}]{volovik1999}%
  \BibitemOpen
  \bibfield  {author} {\bibinfo {author} {\bibfnamefont {G.~E.}\ \bibnamefont
  {Volovik}},\ }\href {\doibase 10.1134/1.568223} {\bibfield  {journal}
  {\bibinfo  {journal} {JETP Lett.}\ }\textbf {\bibinfo {volume} {70}},\
  \bibinfo {pages} {609} (\bibinfo {year} {1999})}\BibitemShut {NoStop}%
\bibitem [{\citenamefont {Simonucci}\ \emph {et~al.}(2015)\citenamefont
  {Simonucci}, \citenamefont {Pieri},\ and\ \citenamefont
  {Strinati}}]{simonucci15}%
  \BibitemOpen
  \bibfield  {author} {\bibinfo {author} {\bibfnamefont {S.}~\bibnamefont
  {Simonucci}}, \bibinfo {author} {\bibfnamefont {P.}~\bibnamefont {Pieri}}, \
  and\ \bibinfo {author} {\bibfnamefont {C.}~\bibnamefont {Strinati}},\ }\href
  {\doibase 10.1038/nphys3449} {\bibfield  {journal} {\bibinfo  {journal}
  {Nature {P}hysics}\ }\textbf {\bibinfo {volume} {11}},\ \bibinfo {pages}
  {941} (\bibinfo {year} {2015})}\BibitemShut {NoStop}%
\bibitem [{\citenamefont {Aghaee}\ \emph {et~al.}(2026)\citenamefont {Aghaee}
  \emph {et~al.}}]{aghaee202620secondparitylifetime}%
  \BibitemOpen
  \bibfield  {author} {\bibinfo {author} {\bibfnamefont {M.}~\bibnamefont
  {Aghaee}} \emph {et~al.},\ }\href {https://arxiv.org/abs/2606.03884}
  {\enquote {\bibinfo {title} {20 second parity lifetime in an inas--pb tetron
  device},}\ } (\bibinfo {year} {2026}),\ \Eprint
  {http://arxiv.org/abs/2606.03884} {arXiv:2606.03884 [cond-mat.mes-hall]}
  \BibitemShut {NoStop}%
\bibitem [{\citenamefont {Zwierlein}\ \emph {et~al.}(2005)\citenamefont
  {Zwierlein}, \citenamefont {Abo-Shaeer}, \citenamefont {Schirotzek},
  \citenamefont {Schunck},\ and\ \citenamefont {Ketterle}}]{zwierlein2005}%
  \BibitemOpen
  \bibfield  {author} {\bibinfo {author} {\bibfnamefont {M.~W.}\ \bibnamefont
  {Zwierlein}}, \bibinfo {author} {\bibfnamefont {J.~R.}\ \bibnamefont
  {Abo-Shaeer}}, \bibinfo {author} {\bibfnamefont {A.}~\bibnamefont
  {Schirotzek}}, \bibinfo {author} {\bibfnamefont {C.~H.}\ \bibnamefont
  {Schunck}}, \ and\ \bibinfo {author} {\bibfnamefont {W.}~\bibnamefont
  {Ketterle}},\ }\href {\doibase 10.1038/nature03858} {\bibfield  {journal}
  {\bibinfo  {journal} {Nature}\ }\textbf {\bibinfo {volume} {435}},\ \bibinfo
  {pages} {1047} (\bibinfo {year} {2005})}\BibitemShut {NoStop}%
\bibitem [{\citenamefont {Bacon}(2006)}]{PhysRevA.73.012340}%
  \BibitemOpen
  \bibfield  {author} {\bibinfo {author} {\bibfnamefont {D.}~\bibnamefont
  {Bacon}},\ }\href {\doibase 10.1103/PhysRevA.73.012340} {\bibfield  {journal}
  {\bibinfo  {journal} {Phys. Rev. A}\ }\textbf {\bibinfo {volume} {73}},\
  \bibinfo {pages} {012340} (\bibinfo {year} {2006})}\BibitemShut {NoStop}%
\bibitem [{\citenamefont {Bombin}(2010)}]{PhysRevA.81.032301}%
  \BibitemOpen
  \bibfield  {author} {\bibinfo {author} {\bibfnamefont {H.}~\bibnamefont
  {Bombin}},\ }\href {\doibase 10.1103/PhysRevA.81.032301} {\bibfield
  {journal} {\bibinfo  {journal} {Phys. Rev. A}\ }\textbf {\bibinfo {volume}
  {81}},\ \bibinfo {pages} {032301} (\bibinfo {year} {2010})}\BibitemShut
  {NoStop}%
\bibitem [{\citenamefont {Dennis}\ \emph {et~al.}(2002)\citenamefont {Dennis},
  \citenamefont {Kitaev}, \citenamefont {Landahl},\ and\ \citenamefont
  {Preskill}}]{Dennis2002}%
  \BibitemOpen
  \bibfield  {author} {\bibinfo {author} {\bibfnamefont {E.}~\bibnamefont
  {Dennis}}, \bibinfo {author} {\bibfnamefont {A.}~\bibnamefont {Kitaev}},
  \bibinfo {author} {\bibfnamefont {A.}~\bibnamefont {Landahl}}, \ and\
  \bibinfo {author} {\bibfnamefont {J.}~\bibnamefont {Preskill}},\ }\href
  {\doibase 10.1063/1.1499754} {\bibfield  {journal} {\bibinfo  {journal} {J.
  Math. Phys.}\ }\textbf {\bibinfo {volume} {43}},\ \bibinfo {pages} {4452}
  (\bibinfo {year} {2002})}\BibitemShut {NoStop}%
\bibitem [{\citenamefont {Kitaev}(1997)}]{Kitaev1997}%
  \BibitemOpen
  \bibfield  {author} {\bibinfo {author} {\bibfnamefont {A.~Y.}\ \bibnamefont
  {Kitaev}},\ }in\ \href {\doibase 10.1007/978-1-4615-5923-8_19} {\emph
  {\bibinfo {booktitle} {Quantum Communication, Computing, and Measurement}}}\
  (\bibinfo  {publisher} {Springer {US}},\ \bibinfo {year} {1997})\ pp.\
  \bibinfo {pages} {181--188}\BibitemShut {NoStop}%
\bibitem [{\citenamefont {Kitaev}(2003)}]{Kitaev2003}%
  \BibitemOpen
  \bibfield  {author} {\bibinfo {author} {\bibfnamefont {A.}~\bibnamefont
  {Kitaev}},\ }\href {\doibase 10.1016/s0003-4916(02)00018-0} {\bibfield
  {journal} {\bibinfo  {journal} {Ann. Phys.}\ }\textbf {\bibinfo {volume}
  {303}},\ \bibinfo {pages} {2} (\bibinfo {year} {2003})}\BibitemShut {NoStop}%
\bibitem [{\citenamefont {Toikka}\ and\ \citenamefont {L\"{a}uchli}()}]{TL}%
  \BibitemOpen
  \bibfield  {author} {\bibinfo {author} {\bibfnamefont {L.}~\bibnamefont
  {Toikka}}\ and\ \bibinfo {author} {\bibfnamefont {A.}~\bibnamefont
  {L\"{a}uchli}},\ }\href@noop {} {\enquote {\bibinfo {title} {Unpublished},}\
  }\BibitemShut {NoStop}%
\bibitem [{\citenamefont {Gottesmann}(1997)}]{Gottesmann:Stabilizer}%
  \BibitemOpen
  \bibfield  {author} {\bibinfo {author} {\bibfnamefont {D.}~\bibnamefont
  {Gottesmann}},\ }\href@noop {} {\  (\bibinfo {year} {1997})}\BibitemShut
  {NoStop}%
\bibitem [{\citenamefont {Poulin}(2005)}]{Poulin:Stabilizer}%
  \BibitemOpen
  \bibfield  {author} {\bibinfo {author} {\bibfnamefont {D.}~\bibnamefont
  {Poulin}},\ }\href {\doibase 10.1103/PhysRevLett.95.230504} {\bibfield
  {journal} {\bibinfo  {journal} {Physical Review Letters}\ }\textbf {\bibinfo
  {volume} {95}},\ \bibinfo {pages} {230504} (\bibinfo {year}
  {2005})}\BibitemShut {NoStop}%
\bibitem [{\citenamefont {Nussinov}\ and\ \citenamefont {{van den
  Brink}}(2015)}]{CompassModelsReview}%
  \BibitemOpen
  \bibfield  {author} {\bibinfo {author} {\bibfnamefont {Z.}~\bibnamefont
  {Nussinov}}\ and\ \bibinfo {author} {\bibfnamefont {J.}~\bibnamefont {{van
  den Brink}}},\ }\href {\doibase 10.1103/RevModPhys.87.1} {\bibfield
  {journal} {\bibinfo  {journal} {Reviews of Modern Physics}\ }\textbf
  {\bibinfo {volume} {87}},\ \bibinfo {pages} {1} (\bibinfo {year}
  {2015})}\BibitemShut {NoStop}%
\bibitem [{\citenamefont {Li}\ \emph {et~al.}(2019)\citenamefont {Li},
  \citenamefont {Miller}, \citenamefont {Newman}, \citenamefont {Wu},\ and\
  \citenamefont {Brown}}]{Li:CompassCodes}%
  \BibitemOpen
  \bibfield  {author} {\bibinfo {author} {\bibfnamefont {M.}~\bibnamefont
  {Li}}, \bibinfo {author} {\bibfnamefont {D.}~\bibnamefont {Miller}}, \bibinfo
  {author} {\bibfnamefont {M.}~\bibnamefont {Newman}}, \bibinfo {author}
  {\bibfnamefont {Y.}~\bibnamefont {Wu}}, \ and\ \bibinfo {author}
  {\bibfnamefont {K.~R.}\ \bibnamefont {Brown}},\ }\href {\doibase
  10.1103/PhysRevX.9.021041} {\bibfield  {journal} {\bibinfo  {journal}
  {Physical Review X}\ }\textbf {\bibinfo {volume} {9}},\ \bibinfo {pages}
  {021041} (\bibinfo {year} {2019})}\BibitemShut {NoStop}%
\bibitem [{\citenamefont {Nussinov}\ and\ \citenamefont
  {Fradkin}(2005)}]{nussinovDiscreteSlidingSymmetries2005}%
  \BibitemOpen
  \bibfield  {author} {\bibinfo {author} {\bibfnamefont {Z.}~\bibnamefont
  {Nussinov}}\ and\ \bibinfo {author} {\bibfnamefont {E.}~\bibnamefont
  {Fradkin}},\ }\href {\doibase 10.1103/PhysRevB.71.195120} {\bibfield
  {journal} {\bibinfo  {journal} {Physical Review B}\ }\textbf {\bibinfo
  {volume} {71}},\ \bibinfo {pages} {195120} (\bibinfo {year}
  {2005})}\BibitemShut {NoStop}%
\bibitem [{ope(2026)}]{open-source_xyz}%
  \BibitemOpen
  \href {https://github.com/laantoi/open-science} {\enquote {\bibinfo {title}
  {{Python code used to compute the group cardinalities and the threshold
  estimates, and to plot the figures}},}\ } (\bibinfo {year}
  {2026})\BibitemShut {NoStop}%
\bibitem [{\citenamefont {Houdayer}\ and\ \citenamefont
  {Hartmann}(2004)}]{PhysRevB.70.014418}%
  \BibitemOpen
  \bibfield  {author} {\bibinfo {author} {\bibfnamefont {J.}~\bibnamefont
  {Houdayer}}\ and\ \bibinfo {author} {\bibfnamefont {A.~K.}\ \bibnamefont
  {Hartmann}},\ }\href {\doibase 10.1103/PhysRevB.70.014418} {\bibfield
  {journal} {\bibinfo  {journal} {Phys. Rev. B}\ }\textbf {\bibinfo {volume}
  {70}},\ \bibinfo {pages} {014418} (\bibinfo {year} {2004})}\BibitemShut
  {NoStop}%
\bibitem [{\citenamefont {Roffe}\ \emph {et~al.}(2020)\citenamefont {Roffe},
  \citenamefont {White}, \citenamefont {Burton},\ and\ \citenamefont
  {Campbell}}]{roffe_decoding_2020}%
  \BibitemOpen
  \bibfield  {author} {\bibinfo {author} {\bibfnamefont {J.}~\bibnamefont
  {Roffe}}, \bibinfo {author} {\bibfnamefont {D.~R.}\ \bibnamefont {White}},
  \bibinfo {author} {\bibfnamefont {S.}~\bibnamefont {Burton}}, \ and\ \bibinfo
  {author} {\bibfnamefont {E.}~\bibnamefont {Campbell}},\ }\href {\doibase
  10.1103/physrevresearch.2.043423} {\bibfield  {journal} {\bibinfo  {journal}
  {Physical Review Research}\ }\textbf {\bibinfo {volume} {2}} (\bibinfo {year}
  {2020}),\ 10.1103/physrevresearch.2.043423}\BibitemShut {NoStop}%
\bibitem [{\citenamefont {Roffe}(2022)}]{Roffe_LDPC_Python_tools_2022}%
  \BibitemOpen
  \bibfield  {author} {\bibinfo {author} {\bibfnamefont {J.}~\bibnamefont
  {Roffe}},\ }\href {https://pypi.org/project/ldpc/} {\enquote {\bibinfo
  {title} {{LDPC: Python tools for low density parity check codes}},}\ }
  (\bibinfo {year} {2022})\BibitemShut {NoStop}%
\bibitem [{\citenamefont {Wang}\ \emph
  {et~al.}(2003{\natexlab{b}})\citenamefont {Wang}, \citenamefont
  {Harrington},\ and\ \citenamefont {Preskill}}]{Wang2003}%
  \BibitemOpen
  \bibfield  {author} {\bibinfo {author} {\bibfnamefont {C.}~\bibnamefont
  {Wang}}, \bibinfo {author} {\bibfnamefont {J.}~\bibnamefont {Harrington}}, \
  and\ \bibinfo {author} {\bibfnamefont {J.}~\bibnamefont {Preskill}},\ }\href
  {\doibase 10.1016/s0003-4916(02)00019-2} {\bibfield  {journal} {\bibinfo
  {journal} {Annals of Physics}\ }\textbf {\bibinfo {volume} {303}},\ \bibinfo
  {pages} {31–58} (\bibinfo {year} {2003}{\natexlab{b}})}\BibitemShut
  {NoStop}%
\bibitem [{\citenamefont {Tuckett}\ \emph {et~al.}(2018)\citenamefont
  {Tuckett}, \citenamefont {Bartlett},\ and\ \citenamefont
  {Flammia}}]{PhysRevLett.120.050505}%
  \BibitemOpen
  \bibfield  {author} {\bibinfo {author} {\bibfnamefont {D.~K.}\ \bibnamefont
  {Tuckett}}, \bibinfo {author} {\bibfnamefont {S.~D.}\ \bibnamefont
  {Bartlett}}, \ and\ \bibinfo {author} {\bibfnamefont {S.~T.}\ \bibnamefont
  {Flammia}},\ }\href {\doibase 10.1103/PhysRevLett.120.050505} {\bibfield
  {journal} {\bibinfo  {journal} {Phys. Rev. Lett.}\ }\textbf {\bibinfo
  {volume} {120}},\ \bibinfo {pages} {050505} (\bibinfo {year}
  {2018})}\BibitemShut {NoStop}%
\bibitem [{\citenamefont
  {Bombin}(2013)}]{https://doi.org/10.48550/arxiv.1311.0277}%
  \BibitemOpen
  \bibfield  {author} {\bibinfo {author} {\bibfnamefont {H.}~\bibnamefont
  {Bombin}},\ }\href {\doibase 10.48550/ARXIV.1311.0277} {\  (\bibinfo {year}
  {2013}),\ 10.48550/ARXIV.1311.0277}\BibitemShut {NoStop}%
\bibitem [{\citenamefont {Gidney}\ \emph {et~al.}(2022)\citenamefont {Gidney},
  \citenamefont {Newman},\ and\ \citenamefont
  {McEwen}}]{Gidney2022benchmarkingplanar}%
  \BibitemOpen
  \bibfield  {author} {\bibinfo {author} {\bibfnamefont {C.}~\bibnamefont
  {Gidney}}, \bibinfo {author} {\bibfnamefont {M.}~\bibnamefont {Newman}}, \
  and\ \bibinfo {author} {\bibfnamefont {M.}~\bibnamefont {McEwen}},\ }\href
  {\doibase 10.22331/q-2022-09-21-813} {\bibfield  {journal} {\bibinfo
  {journal} {{Quantum}}\ }\textbf {\bibinfo {volume} {6}},\ \bibinfo {pages}
  {813} (\bibinfo {year} {2022})}\BibitemShut {NoStop}%
\bibitem [{\citenamefont {Horsman}\ \emph {et~al.}(2012)\citenamefont
  {Horsman}, \citenamefont {Fowler}, \citenamefont {Devitt},\ and\
  \citenamefont {Meter}}]{Horsman2012}%
  \BibitemOpen
  \bibfield  {author} {\bibinfo {author} {\bibfnamefont {D.}~\bibnamefont
  {Horsman}}, \bibinfo {author} {\bibfnamefont {A.~G.}\ \bibnamefont {Fowler}},
  \bibinfo {author} {\bibfnamefont {S.}~\bibnamefont {Devitt}}, \ and\ \bibinfo
  {author} {\bibfnamefont {R.~V.}\ \bibnamefont {Meter}},\ }\href {\doibase
  10.1088/1367-2630/14/12/123011} {\bibfield  {journal} {\bibinfo  {journal}
  {New Journal of Physics}\ }\textbf {\bibinfo {volume} {14}},\ \bibinfo
  {pages} {123011} (\bibinfo {year} {2012})}\BibitemShut {NoStop}%
\bibitem [{\citenamefont {Chatterjee}\ \emph {et~al.}(2025)\citenamefont
  {Chatterjee}, \citenamefont {Das},\ and\ \citenamefont
  {Ghosh}}]{Chatterjee2025}%
  \BibitemOpen
  \bibfield  {author} {\bibinfo {author} {\bibfnamefont {A.}~\bibnamefont
  {Chatterjee}}, \bibinfo {author} {\bibfnamefont {S.}~\bibnamefont {Das}}, \
  and\ \bibinfo {author} {\bibfnamefont {S.}~\bibnamefont {Ghosh}},\ }\href
  {\doibase 10.3390/s25061854} {\bibfield  {journal} {\bibinfo  {journal}
  {Sensors}\ }\textbf {\bibinfo {volume} {25}},\ \bibinfo {pages} {1854}
  (\bibinfo {year} {2025})}\BibitemShut {NoStop}%
\bibitem [{\citenamefont {Anderson}\ \emph {et~al.}(2014)\citenamefont
  {Anderson}, \citenamefont {Duclos-Cianci},\ and\ \citenamefont
  {Poulin}}]{PhysRevLett.113.080501}%
  \BibitemOpen
  \bibfield  {author} {\bibinfo {author} {\bibfnamefont {J.~T.}\ \bibnamefont
  {Anderson}}, \bibinfo {author} {\bibfnamefont {G.}~\bibnamefont
  {Duclos-Cianci}}, \ and\ \bibinfo {author} {\bibfnamefont {D.}~\bibnamefont
  {Poulin}},\ }\href {\doibase 10.1103/PhysRevLett.113.080501} {\bibfield
  {journal} {\bibinfo  {journal} {Phys. Rev. Lett.}\ }\textbf {\bibinfo
  {volume} {113}},\ \bibinfo {pages} {080501} (\bibinfo {year}
  {2014})}\BibitemShut {NoStop}%
\end{thebibliography}%

\end{document}